\documentclass[twocolumn,10pt]{article}
\usepackage[utf8]{inputenc}
\usepackage[T1]{fontenc}
\usepackage{newtxtext,newtxmath}
\usepackage[letterpaper,margin=1in]{geometry}
\usepackage{graphicx}
\usepackage{booktabs}
\usepackage{xcolor}
\usepackage[numbers,sort&compress]{natbib}
\usepackage[hidelinks]{hyperref}
\graphicspath{{./}{../figures/}}
\usepackage{titlesec}
\titleformat*{\section}{\large\bfseries}
\titleformat*{\subsection}{\normalsize\bfseries}
\begin{document}
\title{Who Keeps the Gains from Personal AI Assistants? Seller Adaptation and the Unassisted in a Language-Model Market Simulation}
\author{Haonan Huang\\ Princeton University \and Joey Xiao\\ New York University}
\date{}
\maketitle
\begin{abstract}
Personal AI assistants are beginning to transact for consumers, and early adopters capture real savings. Whether those savings survive, and what happens to consumers who have no assistant, depends on how sellers respond --- a question single-user evidence cannot answer. We build an agent-based rental market in which language models play consumers, assistants, and six adaptive sellers guided by an algorithmic pricing tool. Half the population receives an assistant under an advisory or an executing mandate; the contract pairs a fee only the renter's physical action avoids with a pre-selected add-on an authorised assistant can cancel online. An analytical benchmark and a behaviourally calibrated rule market supply ex-ante predictions, and paired branches with frozen versus adaptive sellers separate adoption effects from market feedback. Across thirty simulated markets, executing assistants cut adopters' spending by 13.7 USD per renter-day when sellers are frozen; adaptation claws back about a third, leaving 8.7, with the gains arriving both as lower bills and as rentals completed at all. Sellers raise headline rates while cutting fees, and the calibrated forecast of the burden on unassisted consumers (+3.6) does not transfer: their mean spending change is +0.4, confidence interval $-$0.6 to +1.3. Seller-model swaps and a within-market transfer of fee-setting to the pricing tool show that fee conduct, and with it the division of the gains, is decided on the seller side. Assistants, we conclude, should be evaluated at market level --- completion, total spending, and non-users included --- and the comparison layers locate exactly where a calibrated behavioural forecast fails in a language-model market.
\end{abstract}

\section{Introduction}

Assistants that act, rather than merely answer, are reaching consumers now: a deployed agent asks its principal before each purchase and operates inside scoped app permissions \cite{meta2026muse}, and the mandate structures that will govern such delegation are being designed in parallel \cite{michael2026permission,kim2026delegation}. Diffusion is already measurably uneven across customers and journey stages \cite{yan2026platformassistant}, so for years every market will mix assisted and unassisted customers. On the individual scale the technology plainly works. In a randomised field marketplace, a shopping tool that surfaces the all-in price eliminated most of the drip-pricing overcharge for its users \cite{lu2025drip}. Two questions decide what that success amounts to socially, and single-user evidence cannot answer either. First, through which channel does the assistant create value: by informing a consumer who would not have read, or by acting where the consumer reads and still fails to act? Second, once sellers whose margins live in contingent fees re-optimise against a partially assisted population, where does the saving end up: with the assisted, with the sellers, or extracted from the bystanders who have no assistant?

Economic theory sharpens the second question into a warning. In models of shrouded attributes, customers who dodge a contingent fee are cross-subsidised by customers who pay it, so any technology that makes avoidance free for one segment pushes the competitive headline price up for everyone \cite{gabaix2006shrouded,armstrong2012contingent}. Assistants would then act as a regressive transfer: adopters save, sellers reprice, bystanders fund the difference. That prediction, however, hinges on behavioural parameters the theory assumes rather than observes, and on sellers who respond only through prices when real sellers also choose disclosure and obstruction.

We answer both questions inside a simulated market small enough to audit and rich enough to bite (Figure 1). A language model (LLM) is first measured, cell by cell, in the roles it will play: a renter living one errand-filled day, and an assistant holding either an advisory or an executing mandate over a single digital act. The measured agents then populate a rental market whose contract carries a daily rate plus two contingent fees engineered for identification: a refuelling fee that only the renter's physical action can avoid, and a pre-ticked add-on whose online cancellation an assistant can perform. Six language-model sellers, each equipped with an algorithmic pricing aid, adapt rates, fees, disclosure placement, and decline-step obstruction. Handing an assistant to every second person, holding a fixed panel of never-adopters aside, and pairing every shocked path with its unshocked twin under common random draws yields three audited quantities: the adopters' outlay change, the bystanders' outlay change, and the sellers' revenue recovery. Design, sample sizes, and reading rules were locked and fingerprinted with no outcome yet in view.

The answers reorder the usual story. Value comes from acting, not reading: the simulated consumer reads the fine print on most days --- hundreds of times field readership --- and still pays, while an assistant holding a concrete, approved plan completes the cancellation almost always (advice nearly matches full delegation) and, more mundanely, makes sure a car is secured at all. Sellers do reprice: they rebuild about half the lost fee stream through the headline rate while abandoning the fees themselves, leaving adopters 8.7 of the 13.7 USD per renter-day they would keep under frozen sellers --- about two-thirds, of which over two-fifths is completed transactions rather than dodged charges. The feared bystander tax, finally, fails to appear --- thirty markets bound it within about a dollar a day --- and what would create it does not belong to the assistant at all; it belongs to the sellers' fee-setting: across seven market configurations spanning two model families, agent-replacement hybrids, and a controlled reassignment of fee-setting rights to the sellers' own pricing tool, the bystander outcome moves from burden to saving in step with the fee margin the market carried when assistants arrived, resolved at the two ends of that scale. Where selling is tool-priced, fees are thin, vanish entirely under the shock, and everyone --- assisted or not --- buys at an all-in price.

\textbf{Contributions.} (1) A measurement-first pipeline in which the same pre-registered estimands pass from an analytical benchmark through a calibrated behavioural market to a language-model society, so that agreement is informative and disagreement is attributable (Sections 3--6). (2) An acting-not-reading account of assistant value, quantified down to its two components --- fee avoidance and task completion --- of which only the first is contestable by sellers (Sections 4, 6). (3) A conduct-dependence result for the distributional question, including a within-market institutional intervention: a controlled transfer of fee-setting to the pricing tool, whose announcement alone makes sellers adopt it pre-emptively and empties the market of fees (Section 7), with the policy reading in Section 8.

\begin{figure*}[t]
\centering
\includegraphics[width=1.00\textwidth]{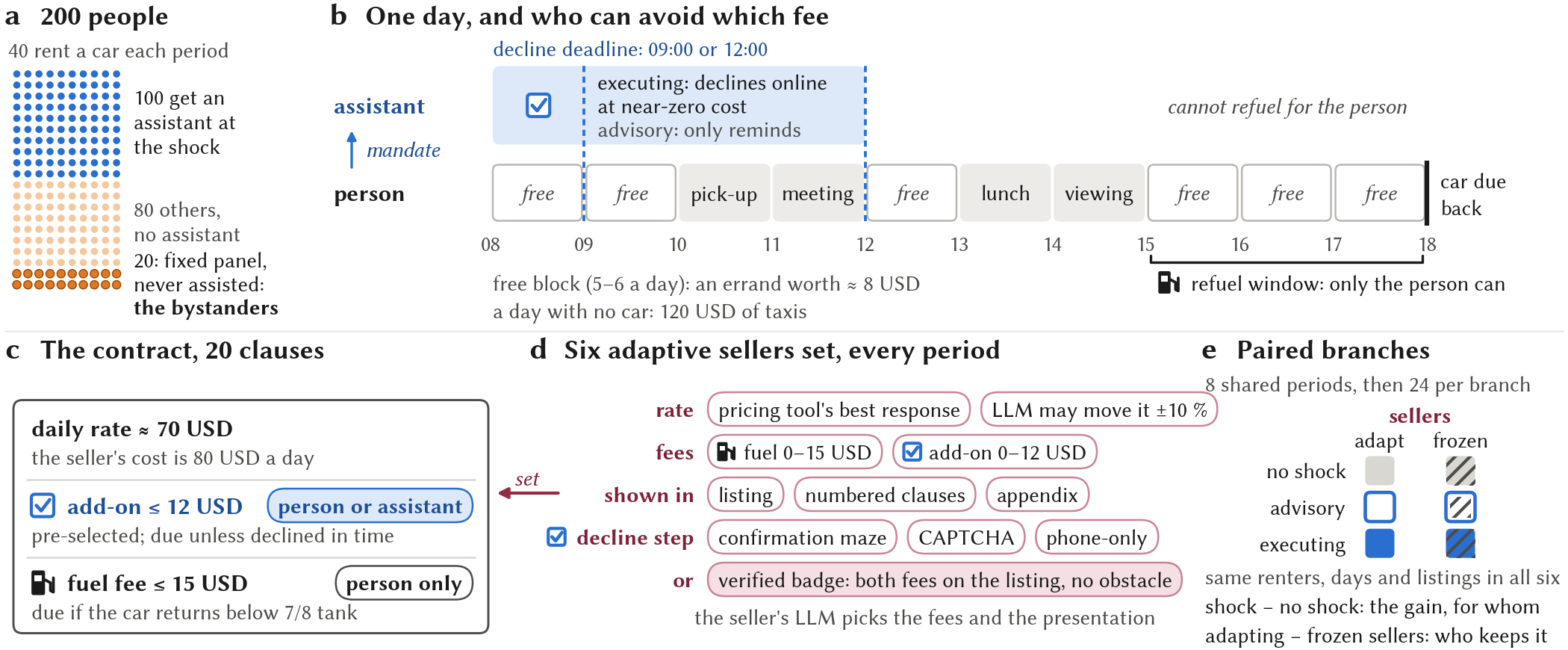}
\caption{The simulation at a glance: (a) two hundred renters, half assisted at the shock, a fixed never-assisted panel --- the bystanders; (b) one day, the mandate's reach, and the refuel window only the renter can use; (c) the contract's rate and two fees, each tied to who can escape it; (d) the sellers' levers, from the pricing tool's rate to obstruction or the badge; (e) the paired branch design that turns differences into the three registered quantities.}
\label{fig:1}
\end{figure*}

\section{Related work}

\textbf{Agentic consumers meeting adaptive sellers.} Theory on AI-mediated purchase finds that cheaper search can help buyers even after sellers respond, while better-informed search can soften competition \cite{lucier2026agentic}; delegating choice to an LLM inside a static duopoly transfers surplus between delegators and non-delegators through a uniform price \cite{delaprez2026welfare}; and simulated airline sellers earn drip revenue only on the human remainder as booking bots spread \cite{abdelghany2026airline}. Reviews of the field flag dynamics and seller-side instrument choice as the open front \cite{rothschild2026evolution}. A cluster of adjacent designs --- delegation depth \cite{want2026delegate}, autonomous search \cite{dong2026agentic}, selling bots against buying bots \cite{xie2026pricing,fu2026strategic}, platform gatekeeping of comparison agents \cite{chen2026open}, collusion under consumer AI \cite{cheng2026collusion}, endogenous sophistication \cite{balzer2026exploitation}, agents with their own attention frictions \cite{bichler2026agentic} --- and the open agent marketplaces \cite{karten2026bazaar,bansal2025magentic} share our motivation but none crosses partial adoption with a fee that only some customers can escape. On the seller side, deployed agents shade truth under incentive \cite{liu2026knownliebench} and drift toward profit framings unprompted \cite{so2026profit}; assistant memory can discipline platforms at full adoption \cite{hong2026platforms}; long-horizon agent economies are unstable \cite{regmi2026town,li2026misaligned}, which our short, zero-sum-audited window avoids.

\textbf{Contingent charges.} The cross-subsidy logic we test descends from add-on pricing \cite{ellison2005addon}, shrouding equilibria \cite{gabaix2006shrouded}, and contingent-charge contracts \cite{armstrong2012contingent}, with the survival of exploitative terms under competition \cite{heidhues2016exploitative,heidhues2017inferior}, obfuscation as strategy \cite{spiegler2006competition,spiegler2016complexity}. Empirically, banned fees migrate into salient prices at most partially \cite{agarwal2015regulating} --- our recovery benchmark --- and shrouding moves demand \cite{brown2010shrouded,blake2021salience}; mandated all-in pricing shifts, rather than destroys, fee revenue \cite{yang2026drip,jin2026clearing}. The assistant-facing frictions we model are documented practice: incumbent suits against buying agents \cite{vanloo2026consumer}, bot walls and CAPTCHAs \cite{bandara2026agentfirst}, and dark-pattern incidence benchmarks \cite{ersoy2025darkpatterns,tang2025darkpatterns,shi2026webdecept,ousat2026brokengates}.

\textbf{Simulating with language models.} LLM agents reproduce some economic behaviour \cite{horton2023homo,argyle2023outofone,aher2023simulate,park2023generative} and miss other parts --- competitive convergence \cite{struski2026competitive}, pre-registered treatment effects \cite{tsai2026seeing}, outside options \cite{chen2026simulated}. We adopt the construct-validity discipline urged for such simulations \cite{zhu2026guardrails}: measure the agent first, predict with a transparent baseline, register the read, and report the simulator's own noise. Benchmarks of shopping-agent behaviour \cite{wadi2026rank,wadi2026shopping} anchor our assistant-side measurements.

\section{A market that fits in one day}

This section walks the construction of Figure 1: the day, the contract, the three parties, and the design; the supplement lists every constant.

\begin{figure*}[t]
\centering
\includegraphics[width=1.00\textwidth]{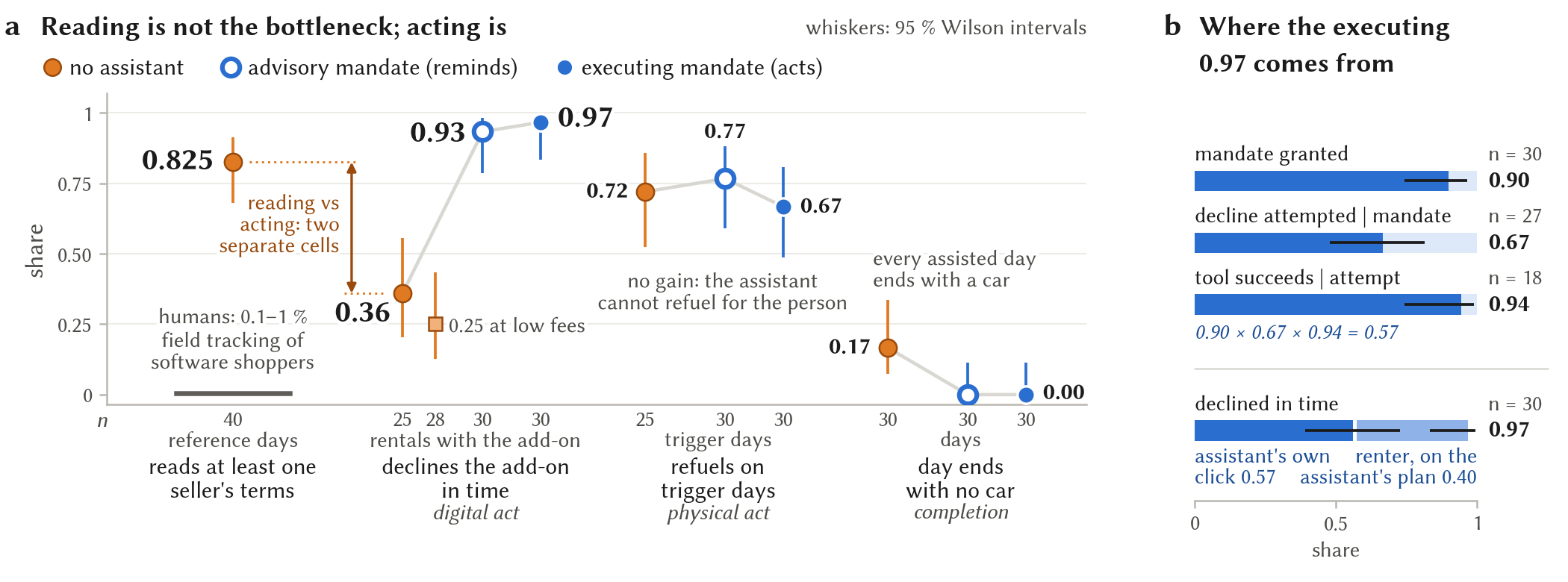}
\caption{Reading against acting in the role measurements: (a) shares of days or rentals on which the renter reads, cancels in time, refuels, and ends carless --- unassisted against both mandates, 95 \% Wilson intervals, denominators beneath each cell, the 0.1--1 \% human readership band from field tracking \cite{bakos2014fineprint}; (b) the executing chain's leak points and the split of timely cancellations between the assistant's own click and the renter acting on its plan.}
\label{fig:2}
\end{figure*}

\textbf{One renter-day.} Each simulated person lives ten hourly blocks between 08:00 and 18:00, holding four fixed appointments and five or six open hours; each open hour can absorb one errand worth about 8 USD. Errands require the rental car, which must come back by 18:00. A day on which no car gets secured falls back to taxis, billed at 120 USD. The renter carries a scratchpad of bounded length, picks one action per open hour from a menu of currently feasible moves, and is told nothing else: no fee, no deadline, and no reminder ever appears in its instructions, so any avoidance it achieves is its own.

\textbf{The contract.} Six sellers list a daily rate near 70 USD against a marginal cost of 80 per rented day, recouping the gap through two contingent charges buried in a twenty-clause agreement. A \emph{refuelling charge}, capped at 15 USD, falls due when the tank returns below seven-eighths; escaping it requires the renter itself to spend an open hour at the pump after its last drive --- no software can do this. An \emph{add-on bundle}, capped at 12 USD, arrives pre-ticked and bills automatically unless cancelled ahead of a deadline stated deep in the agreement; the cancellation itself is a two-minute online act. The pair is the identification device: one charge reachable by delegation, one not.

\textbf{The assistant.} Half the population, drawn once and stratified on measured reading propensity, receives an assistant --- the identical model given a clean context and authority over every seller's small print. Holding only the \emph{advisory} mandate it summarises, recommends, and sets reminders, while the renter still clicks. Promoted to the \emph{executing} mandate, it additionally carries out the cancellation itself, once per rental, after its plan wins the renter's approval. It also books the renter's lunch under both mandates, so mandate contrasts isolate the cancellation right. A tenth of the population is a permanently unassisted panel --- our bystanders --- and the remainder without assistants complete the demand side.

\textbf{The sellers.} Every period each seller's LLM receives its ledger and its \emph{pricing tool's} recommendation --- a best-response rate and fee schedule computed from the seller's own observed paying rates at the measured demand elasticity --- and chooses: both fee levels; where each charge appears (listing card, numbered clause, or appendix); whether to obstruct the cancellation path with a phone-only requirement, a CAPTCHA, or a confirmation maze; whether to qualify for a verified transparency badge by listing both charges and obstructing nothing; and a headline rate within ten percent of the tool's quote for the chosen package. Sellers never observe who holds an assistant; a mishandled charge costs them an 8 USD complaint. Obstruction is not free theatre: a phone-only path defeats assisted cancellation almost surely, and each obstacle carries a measured or declared failure probability for renter and assistant alike.

\textbf{Design and estimands.} A market instance is burned in for eight periods with no assistants, snapshotted, then branched for twenty-four periods: an unshocked continuation, the two mandate shocks, and copies of each with sellers frozen at the snapshot --- six branches sharing renters, days, and draws through common random numbers, so differences are treatment, not luck. The final eight periods supply the three registered quantities, market by market, the market being the inference unit; forty of the two hundred rent in any period, hence \emph{per sampled renter-day}. Prose quotes savings and burdens as positive magnitudes; figures plot the signed outlay change. The \emph{adopters' saving}: spending per sampled renter-day among would-be adopters, shocked minus unshocked, carless days entering at the taxi fallback. The \emph{bystanders' burden}: the same difference for the fixed panel. The \emph{sellers' recovery}: revenue regained by adaptive sellers relative to frozen ones, as a fraction of the fee revenue freezing shows they lost, split into a rate part and a fee part. A twin unshocked run supplies the simulator's own replication noise; any contrast whose interval overlaps that scale is reported as unresolved, not as a finding. Estimands, market counts, windows, and these reading rules were frozen and fingerprinted ahead of the first branch being opened; the pilot-stage interface repairs and each later addition carry dated amendments in the released log.

\section{Why assistance works: measured, not assumed}

The renter reads the terms on 0.825 of days yet beats the cancellation deadline on only 0.36 of rentals: that pair of numbers, measured before any market ran, is the paper's first result. The role measurements behind it use acceptance bands set beforehand and thirty-to-forty-day cells, each rate carrying its own denominator. Renters and assistants run on Qwen3.8-27B in four-bit quantisation and sellers on Gemma-4-31B; a sixteen-bit check reproduces the renter cells within their intervals; no proprietary API touches any run, and prompts ship with the release.

\textbf{Information is not the constraint.} The simulated renter opens somebody's fine print on 0.825 of reference days --- two to three orders of magnitude above the 0.1--1 \% of shoppers whom field tracking catches opening a licence \cite{bakos2014fineprint} --- and its reading barely responds to workload. Acting is the constraint: at the reference fee its cancellation beats the deadline on 0.36 of rentals (0.25 when little is at stake), and it refuels on 0.72 of the days the tank rule binds. The screening gate we had registered treated heavy reading as disqualifying; the model failed that gate on both families, and we proceeded under the registered fallback --- the population is \emph{readers who under-execute}, and every market claim is a claim about that population. Its price sensitivity, a conditional-logit scale of 0.84 USD from rate-dispersed listings, feeds the sellers' pricing aid, so seller beliefs match measured demand by construction.

\textbf{Acting is.} Figure 2 shows the flip. Given the executing mandate, the assistant lifts timely cancellation from 0.36 to 0.97; advice alone reaches 0.93, because this renter, unlike its unassisted baseline, follows a concrete plan it has approved. The physical charge moves the other way: refuelling on threatened days edges from 0.72 down to 0.67 under execution --- delegating the digital chore slightly crowds out the pump. And a channel invisible to fee accounting appears: one in six unassisted renter-days ends with no car secured at all, against none with an assistant of either mandate. The mechanism is mundane. An unassisted renter can burn its free hours without ever finishing a booking, the more often the higher the posted fees (0.07 carless at the low fee, 0.17 at the reference); the channel therefore mixes price response with pure follow-through, and an assistant restores the follow-through. Where the executing chain leaks, it leaks at authority and initiative: the mandate arrives on 0.90 of bookings, an attempt follows on two-thirds of grants, the cancellation tool succeeding on 0.94 of attempts. Only 0.57 of the timely cancellations are the assistant's own click; the rest are the renter carrying out the plan it approved. Against seller obstruction the asymmetry sharpens: a phone-only cancellation path caps assisted timely success at 0.13, while the verified transparency badge is worth near 2 USD to the assistant and near 6.5 to the renter.

\section{What repricing should do: benchmark and calibrated forecast}

Two transparent layers, run before the society and frozen as its forecast, say what seller adaptation ought to look like if the measurements are taken seriously.

\textbf{A pencil-and-paper bound.} In a zero-profit contract with a contingent charge, the customers who pay it subsidise the headline for those who dodge it \cite{armstrong2012contingent}. Remove the add-on stream from half the customer base and the break-even headline must climb by the per-customer fee margin times the adopting share --- 5 USD here at half adoption --- while the charge stays profitable on whoever remains. That is the cross-subsidy warning in its purest form: bystanders pay more, assisted or not. The bound also announces its own fragility: at the measured behaviour, a fee-free entrant already beats the incumbent contract before any assistant exists, so the fee-laden configuration is a repricing baseline rather than a resting point.

\textbf{A market of the measured parameters.} Substituting every measured rate into deterministic renters, assistants, and tool-following sellers --- no language model in the loop --- and running the identical branch design 120 times gives the \emph{rule market}, the calibrated forecast the society is judged against. Its adapted state is fee-light and dearer per day: under the executing shock the headline climbs 9.6 USD, adopters retain about two-fifths of a 13.8 USD first-day saving, recovery lands at 0.66 of the lost fee stream --- the migration fraction observed after real fee regulation \cite{agarwal2015regulating} --- and the bystander burden settles at 3.60 $\pm$ 0.34 per renter-day. The forecast, frozen there, will prove right in direction and high on pass-through. One of its ingredients deserves emphasis: bystander burden flips sign when sellers' pricing beliefs loosen, so the forecast's own sensitivity analysis already nominates seller conduct, not assistant capability, as the decisive unknown.

\begin{figure*}[t]
\centering
\includegraphics[width=1.00\textwidth]{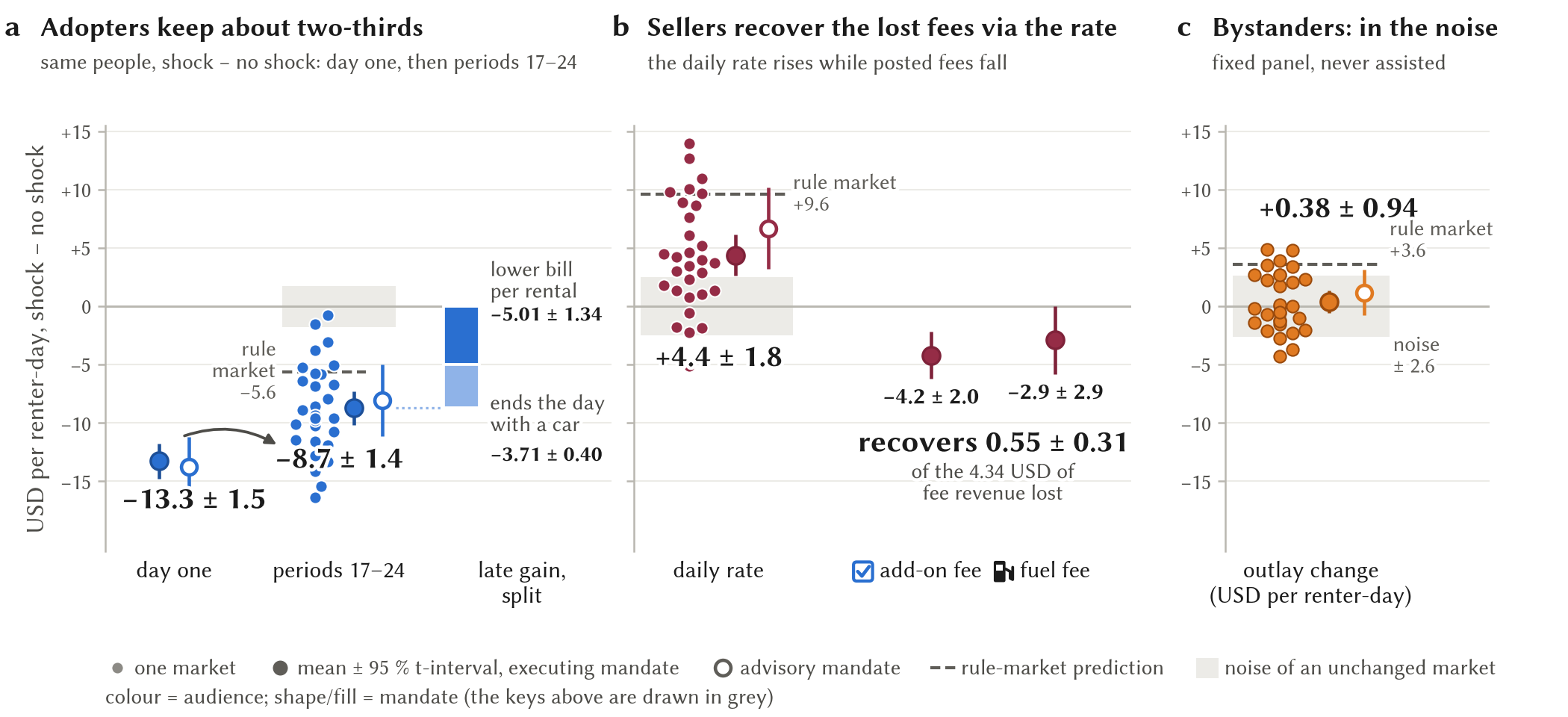}
\caption{The shock's three audiences over thirty markets, small dots one market each, filled marks executing and hollow advisory (advisory arms exist on the registered ten): (a) adopters on day one and adapted, with the adapted saving split into the cheaper bill and cars secured; (b) the sellers' side, rate change beside the recovery of the lost fees; (c) bystanders against the rule-market forecast. The grey band is the twin-run scale of a single unchanged market, three pairs, printed for calibration only.}
\label{fig:3}
\end{figure*}

\section{What the market did}

\textbf{Adopters keep about two-thirds, in two currencies.} Thirty market instances ran the branch design with LLM renters, assistants, and sellers --- the ten registered up front, then twenty more under a replication protocol declared before launch, pooled here as that protocol fixed and read once by the registered script (Figure 3, Table 1). On the first shocked day, would-be adopters spend 13.3 $\pm$ 1.5 USD per renter-day less than their unshocked selves; by the final eight periods, 8.7 $\pm$ 1.4 of it remains --- 66 \%, where the rule market forecast two-fifths. An accounting identity --- outlay equals the carless share times the taxi cost plus the rest times the bill --- splits the surviving saving into 5.0 USD through a cheaper bill and 3.7 USD through cars secured, as carless days collapse from 0.148 to 0.002 of adopter renter-days. Only the cheaper-bill part is fee arbitrage; the cars-secured part is follow-through, and the seller responses below barely reach it.

\textbf{Sellers rebuild half the lost fees through the rate while abandoning the instrument.} Freezing sellers shows the shock costs them 4.3 USD of fee revenue per rental; adaptive sellers regain 0.55 $\pm$ 0.31 of it, all through a headline rate 4.4 $\pm$ 1.8 USD higher (rate component 1.20, fee component $-$0.65), under both the forecast 0.66 and the migration fraction real fee regulation produced. The instrument itself is dying where it is being repriced: both posted fees fall, one seller-period in five goes fee-free against almost none unshocked, and the \emph{unassisted} get billed the add-on on 30 points fewer rentals --- a fee a seller stops charging stops being charged to everyone. Transparency, for its part, belongs to the reading population rather than to the shock: 5.6 of six sellers hold the badge, a third of a seller obstructs, and prices keep cycling at 2.1 USD per period into the final window, an adaptation transient honestly averaged, not an equilibrium.

\textbf{Bystanders: no detectable burden.} Under the executing shock the fixed panel's spending shifts by +0.38 $\pm$ 0.94 USD per renter-day over the thirty markets --- zero inside the interval, the whole effect bounded within about $\pm$1.3 on a 100 USD daily outlay, and +0.13 $\pm$ 0.73 over the whole horizon. The declared protocol reports its three cuts side by side: the registered ten alone read +1.71 $\pm$ 1.61, the twenty replication markets $-$0.28 $\pm$ 1.14, the pool as above; the twin-run noise scale of a single market, 2.6 USD, dwarfs all three. The replication also answers the registered ten's one oddity --- their five viability flags and stronger repricing: all twenty new markets are viable, their sellers reprice less (rate +3.1 $\pm$ 2.2, recovery 0.33 $\pm$ 0.43), and the pooled society accordingly sits \emph{below} the rule-market forecast on pass-through while its adopters keep more. With thirty markets the society's precision separates it from its forecast: the bystander, rate, and adopter intervals now exclude the rule market's means while recovery covers --- the deterministic layer called the direction of every response and overstated how much of it sellers complete. The advisory mandate, run on the registered ten, tells the stronger-for-adopters story there ($-$8.1 $\pm$ 3.1 against the chain's $-$4.2) with the same null bystander reading.

\begin{table}[t]
\caption{Rule-market forecast against society outcome, executing mandate, final eight periods; USD per sampled renter-day except rows marked as shares or ratios. Society: thirty pooled markets (the registered ten and the twenty-market replication are reported separately in the text; advisory column in the released materials).}
\footnotesize
\setlength{\tabcolsep}{4pt}
\begin{tabular}{p{0.40\columnwidth}rrr}
\toprule
registered quantity & forecast & outcome (30) & twin-run noise \\
\midrule
adopters' saving, first day & 13.8 $\pm$ 0.4 & 13.3 $\pm$ 1.5 & --- \\
adopters' saving, adapted & 5.6 $\pm$ 0.3 & 8.7 $\pm$ 1.4 & 1.8 \\
\ldots{} of it through cars secured (share) & --- & 0.43 & --- \\
bystanders' burden, adapted & 3.60 $\pm$ 0.34 & 0.38 $\pm$ 0.94 & 2.6 \\
headline rate rise & 9.6 $\pm$ 0.3 & 4.4 $\pm$ 1.8 & 2.5 \\
sellers' recovery of lost fees (ratio) & 0.66 $\pm$ 0.03 & 0.55 $\pm$ 0.31 & --- \\
\ldots{} through rate / through fees (ratio) & 0.81 / $-$0.15 & 1.20 / $-$0.65 & --- \\
adopters' add-on billed less often (share of rentals) & 0.99 & 0.74 $\pm$ 0.03 & --- \\
fee-free seller-periods under the shock (share) & --- & 0.19 & --- \\
\bottomrule
\end{tabular}
\end{table}

\begin{figure*}[t]
\centering
\includegraphics[width=1.00\textwidth]{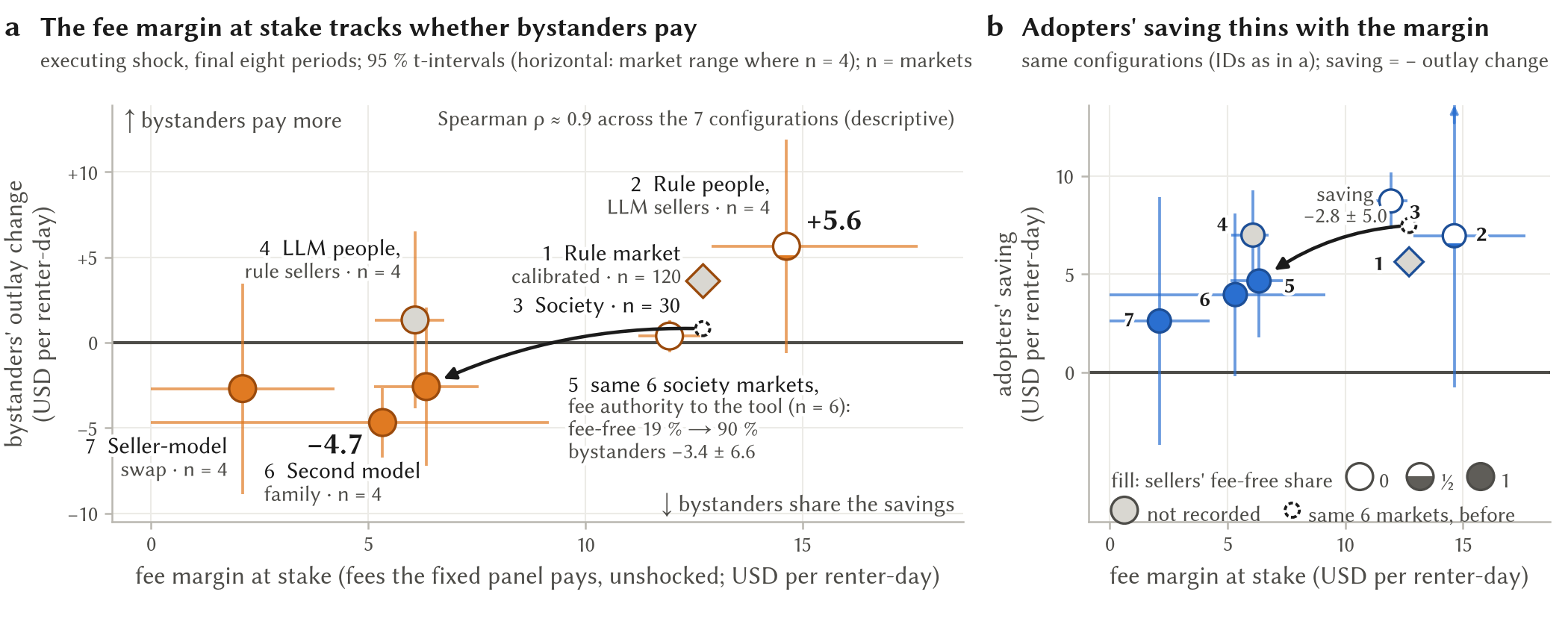}
\caption{Seven configurations on one axis --- the fee margin at stake, the fees the fixed panel pays unshocked: (a) the bystander outcome crosses from burden to saving as the margin thins, resolved at the ends; the dashed ring is the six-market baseline of the fee-authority arm and the arrow its within-market shift, fee-free seller-periods roughly 0.2 to 0.9; (b) the adopters' saving, positive up, thins as the margin does. Final eight periods; intervals on both axes.}
\label{fig:4}
\end{figure*}

\section{Seller conduct decides who pays}

The society's null bystander reading could mean assistants are benign, or could mean this market had little left to extract. Seven configurations separate the two: reruns of the same environment varying which tier is an LLM, which model family plays it, and who controls fees. Figure 4 aligns them on one axis --- the fee margin at stake, measured as what the fixed panel pays in fees per renter-day in each configuration's unshocked state.

\textbf{The margin orders the outcome, and the margin is conduct's footprint.} At the fee-heavy end, the rule market (12.7 USD at stake) taxes bystanders a resolved 3.6, and LLM sellers facing non-reading deterministic renters (14.6 at stake) tax them 5.6 while obstructing half the storefronts. At the fee-light end, three configurations whose sellers hew to the pricing tool --- a second model family in every role, a swap of the seller model alone, and the society under transferred fee authority --- carry 2 to 6 USD at stake, answer the shock with little or no rate rise ($-$0.7 to +3.5), and show bystanders \emph{saving}: $-$4.7 $\pm$ 2.0, $-$2.7 $\pm$ 6.2, $-$2.6 $\pm$ 4.6. Only the two ends are resolved --- the rule market's burden, the second family's saving --- and the rest carry intervals through zero; adopters' saving thins with the margin, its completed-transactions component the slower-moving part. Across the seven configurations the association is near-monotone (Spearman about 0.9), a descriptive synthesis across designed variants, pattern rather than test. Conduct also acts beyond the margin. At the same 12.7 at stake, the society's sellers --- who take the tool's fee advice on a quarter of their choices --- pass through a tenth of what the fully strategic rule market does (+0.4 against +3.6). And where deterministic tool-followers meet reading LLM renters, half the fees are gone before any assistant arrives; the residual burden (+1.3 $\pm$ 5.2) is already indistinguishable from zero.

\textbf{The seller model carries its conduct.} Family comparisons braid three changes together. One arm therefore changes the seller model alone: same renters, same assistants, same seeds, sellers swapped from the fee-defending family to the tool-following one. The swapped market reproduces the second family wholesale --- 87 \% fee-free seller-periods, universal badges, zero obstruction, flat rates --- establishing that fee conduct travels with the seller model, not with the population it faces.

\textbf{Announcing tool-set fees empties the market.} The sharpest probe transfers fee-setting to the pricing tool inside six of the society's own markets, everything else pinned. Sellers are told the tool's fee schedule will be executed; their LLMs keep presentation, obstruction, badges, and bounded rate discretion. Sellers do not resist; they anticipate, proposing the tool's own schedule on 99 \% of seller-periods, and fee-free seller-periods jump from 0.19 to 0.89 of the total. Read as the announced institution it is, window by window. Over the whole horizon adopters' saving halves: 10.4 falling to 5.1 per renter-day, a 5.3 $\pm$ 4.2 USD drop with every one of the six markets positive; the registered final window puts the drop at 2.8 $\pm$ 5.0. About 29 \% of the whole-horizon change flows through the cars-secured channel, because fee-free sellers rescue marginal \emph{unshocked} rentals too, trimming even the follow-through slice. Bystanders' whole-horizon cell goes $-$0.7 to $-$3.0 (a $-$2.3 $\pm$ 4.3 change) and the final-window cell +0.8 to $-$2.6 ($-$3.4 $\pm$ 6.6): the tool-following direction, both spans crossing zero, the latter leaning on one market. The institutional claim therefore stays with the resolved cells. Fees vanish, the contestable half of assistant value vanishes with them, and nothing indicates bystanders are hurt in the process.

\begin{table}[t]
\caption{The fee-authority transfer over six paired markets, in USD for each sampled renter-day, split by window (per-market detail in the released materials).}
\footnotesize
\setlength{\tabcolsep}{4pt}
\begin{tabular}{p{0.40\columnwidth}rrr}
\toprule
quantity $\cdot$ window & sellers set fees & tool sets fees & change \\
\midrule
unshocked renters' outlay $\cdot$ whole & 101.3 & 99.8 & $-$1.5 \\
assisted renters' outlay $\cdot$ whole & 90.9 & 94.7 & +3.8 \\
adopters' saving $\cdot$ whole & 10.4 & 5.1 & $-$5.3 $\pm$ 4.2 \\
adopters' saving $\cdot$ final eight & 7.5 & 4.7 & $-$2.8 $\pm$ 5.0 \\
bystanders' burden $\cdot$ whole & $-$0.7 & $-$3.0 & $-$2.3 $\pm$ 4.3 \\
bystanders' burden $\cdot$ final eight & +0.8 & $-$2.6 & $-$3.4 $\pm$ 6.6 \\
fee-free seller-periods (share) $\cdot$ shock & 0.19 & 0.89 & +0.70 \\
\bottomrule
\end{tabular}
\end{table}

\section{What it means}

\textbf{For the junk-fee debate.} Regulators pursue all-in pricing by rule \cite{ftc2025fees,yang2026drip}; in this market something close to it emerged from automation on both sides of the counter --- an outcome of these simulations, offered as a hypothesis about real markets rather than a conclusion. Executing assistants operate as decentralised enforcement, since every cancellation they complete taxes the shrouding strategy. Tool-driven fee-setting on the seller side finishes the job: a pricing tool computing on observed paying rates simply stops recommending a charge its market has learned to dodge. The convergence point, reached here without a statute, is a dearer honest headline and no contingent charges: the CARD-Act pattern of partial fee-to-rate migration \cite{agarwal2015regulating}, run to completion. The equity objection to consumer AI --- that it prices the unprotected out --- appears in our data only where fee margins are thick and defended. In the society it is smaller than the simulator's own noise, and it inverts where selling is tool-priced. What deserves the regulatory eye instead is the cancellation step. A phone-only requirement neutralises assistance almost completely in our measurements; obstruction of agents, already visible in the field \cite{vanloo2026consumer,bandara2026agentfirst}, is the one seller instrument that re-segments a market assistants would otherwise clean.

\textbf{For the assistant business.} Half of the value that survives repricing here is fee arbitrage (on day one, closer to three-quarters), and that share is self-liquidating: sellers reprice it away for adopters, and competition or institutions erase it for everyone. The part that persisted in every configuration we ran is follow-through --- the assistant completes the transaction its principal fumbles --- worth 3.75 $\pm$ 0.64 USD per renter-day within the society, and dented only where fee-free selling rescues the unshocked side's rentals as well. Products sold on fee savings price their own obsolescence; products built on execution reliability sell what the market left standing.

\textbf{Validity, honestly.} We follow the guardrails literature for LLM simulations \cite{zhu2026guardrails}: agents measured before deployment, forecasts frozen before outcomes, estimands and rules hashed in advance, the simulator's replication noise printed beside every primary contrast. Convergence gets reported, never presumed \cite{struski2026competitive}. The registered screening gate failed --- the model reads too much --- and we proceeded under the registered fallback. External validity is therefore explicitly conditional: this is a market of tireless readers who under-execute, one plausible corner of the coming assisted economy, not a calibrated portrait of human renters. Seller conduct findings are configuration contrasts over a handful of model families; the margin-burden alignment of Figure 4 is descriptive; the fee-authority transfer identifies an announced institution, not a concealed counterfactual; and a twenty-market replication of the society, registered before launch, was still running at submission with its analysis frozen. The carless-day channel earns its own caution: it is an LLM's follow-through failure priced at the taxi fallback, and though it responds to fees as human non-booking would, its level need not transfer to people.

\textbf{The sentence to keep.} Assistants help because they act; markets answer by folding fees into honest prices; who funds the transition is written in the sellers' fee ledger, not in the assistant's code. And whenever either side automated the fee decision here, the fees, the burden, and the arbitrage shrank together, leaving completed transactions as the value still standing.

\bibliographystyle{unsrtnat}
\bibliography{refs}
\end{document}